\documentclass[aps,prd,preprintnumbers,nofootinbib,preprint]{revtex4}
\usepackage{bm}
\usepackage{latexsym}
\usepackage{dcolumn}
\usepackage{amsmath,amsfonts,amssymb}
\usepackage{graphicx,epsfig}
\usepackage{amsthm}
\usepackage{url,hyperref}

\usepackage{slashed}
\newcommand{\be}{\begin{eqnarray}}
\newcommand{\ee}{\end{eqnarray}}
\newcommand{\bea}{\begin{eqnarray}}
\newcommand{\eea}{\end{eqnarray}}

\def\comment#1{}

\usepackage{color}

\definecolor{darkred}{rgb}{.8,0,0}

\definecolor{darkblue}{rgb}{0,0,.7}

\definecolor{darkgreen}{rgb}{0,.7,0}

\begin{document}

%
%
%%%%%%%%%%%%%%%%%%%%%%%%%%%%%%%%%%%%%%%%%%%%%%%%%%%%%%%%%%%%%%
\title{Modified Fermions Tunneling Radiation from Non-stationary, Axially Symmetric Kerr Black Hole}
%%%%%%%%%%%%%%%%%%%%%%%%%%%%%%%%%%%%%%%%%%%%%%%%%%%%%%%%%%%%%%
%
%
%
%
%
\author{{Jin Pu}$^{1,2}$}\email[email:~]{pujin@cwnu.edu.cn}
\author{Kai Lin$^{3}$} \email[email:~]{lk314159@hotmail.com}
\author{Xiao-Tao Zu$^{1}$\vspace{1ex}}\email[email:~]{xtzu@uestc.edu.cn}
\author{Shu-Zheng Yang$^{2}$}\email[email:~]{szyangphys@126.com}
\affiliation{$^1$School of Physics, University of Electronic Science and Technology of China, Chengdu 610054, China\vspace{1ex}}
\affiliation{$^2$College of Physics and Space Science, China West Normal University, Nanchong 637002, China\vspace{1ex}}
\affiliation{$^3$Hubei Subsurface Multi-scale Imaging Key Laboratory, Institute of Geophysics and Geomatics, China University of Geosciences, Wuhan 430074, Hubei, China\vspace{1ex}}
%
%
%
%
%\date{\today}
%
%%%%%%%%%%%%
%%%%%%%%%%%%
\begin{abstract}
%%%%%%%%%%%%
%
%
%
%
%
%
\par\noindent
In this paper, by applying the deformed dispersion relation in quantum gravity theory, we study the correction of fermions' tunneling radiation from non-stationary symmetric black holes. Firstly, the motion equation of fermions is modified in the gravitational spacetime. Based on the motion equation, the modified Hamilton-Jacobi equation has been obtained by a semiclassical approximation method. Then, the tunneling behavior of fermions at the event horizon of non-stationary symmetric Kerr black hole is investigated. Finally, the results show that in the non-stationary symmetric background, the correction of Hawking temperature and the tunneling rate are closely related to the angular parameters of the horizon of the black hole background.\\
\end{abstract}
%
%\pacs{03.65.Ta, 03.65-w}
%
%
%
\maketitle
%
%
%
%
%
%
%%%%%%%%%%%%%%%
\section{Introduction}
\label{intro}
%%%%%%%%%%%%%%%
%
%
\par\noindent
Since Hawking proposed that black holes can radiate thermally like a black body in 1974 \cite{1,2}, a series of studies have been carried out on static, stationary and non-stationary black holes. Actually, Hawking thermal radiation is a pure thermal radiation, and the radiation spectrum formed by this radiation is a pure thermal radiation spectrum, which leads to the problem of the information loss of black holes. In order to explain the information loss paradox of black holes, Robinson, Wilczek, Kraus and Parikh have modified Hawking pure thermal radiation spectrum, and found that the information was conserved during Hawking tunneling radiation from static and stationary black holes, by considering the self-gravitational interaction and the change of curved spacetime background \cite{3,4,5,6,7,8}. Subsequently, there are a series of studies on the massive particles and fermions via tunneling radiation from black holes \cite{9,10,11,12,13,14,15,16,17,18,19,20,21,22,23,24,25,26,27,28,29,30,301,302,303,304,305,306}. However, the actual existence of black holes in the universe should be non-stationary, so the issues, such as the thermodynamic properties and the information conservation of non-stationary black holes, and merging process of black holes, deserve to be studied in depth.
\par
On the other hand, quantum gravity theory suggest that Lorentz symmetry may be modified at high energy. Although the dispersion relation theory at high energy has not yet been fully established, it is generally accepted that the scale of this correction term is equal to or close to the Planck scale. We have studied fermions' quantum tunneling radiation from stationary black holes by using the deformed dispersion relation, and obtained the very interesting results that there was a correction in the tunneling radiation behavior \cite{31}. However, the correction is only obtain for fermions' tunneling radiation at the event horizon of stationary black holes. Therefore, we generalize the modified dispersion relation to study the quantum tunneling radiation of non-stationary symmetric Kerr black holes in this paper, and give an effective correction of thermodynamic characteristics of the black holes.
\par
The remainders of this paper are outlined as follows. In Sec.\ref{2}, by applying the modified dispersion relation in quantum gravity theory, we construct new Rarita-Schwinger equation, and obtain the modified Hamilton-Jacobi equation of fermions by using semiclassical approximation method. In Sec.\ref{3}, the quantum tunneling radiation of fermions from non-stationary symmetric Kerr black hole is modified correctly, and the tunneling rate and Hawking temperature is modified. Sec.\ref{4} ends up with some discussions and conclusions.
%
%
%%%%%%%%%%%%%%%%%%%%%%%%%%%%%%%
\section{The modified Hamilton-Jacobi equation}
\label{2}
%%%%%%%%%%%%%%%%%%%%%%%%%%%%%%%
%
%
%
\par\noindent
Kerner and Mann studied the quantum tunneling radiation of the Dirac field using a semiclassical method \cite{16}. Subsequently, this method was extended to study quantum tunneling radiation in various black holes. Since the kinematic equation of fermions is the Dirac equation, but the Dirac equation is related to the matrix equation, so a new method is proposed in the literature \cite{17} to study the tunneling radiation of Dirac particles in curved spacetime of static and stationary black holes. This method is that, the Dirac equation is transformed into a simple matrix equation, and then this matrix equation is converted into the Hamilton-Jacobi equation in the curved spacetime by applying the relationship between the gamma matrix and the space-time metric. After this Hamilton-Jacobi equation was proposed in 2009, it not only promoted the study of quantum tunneling radiation of dynamic black holes, but also effectively simplified the research work on quantum tunneling radiation of fermions. Recently, this Hamilton-Jacobi equation, combined with the modified Lorentz dispersion relation, has been generalized to effectively revise the quantum tunneling radiation of fermions from the event horizon of stationary axisymmetric Kerr-Nemman de sitter black hole, and has obtained very meaningful results \cite{31}. However, it only modifies the quantum tunneling radiation of fermions from the stationary black hole, while the real black holes existing in the universe are non-stationary, so the quantum tunneling radiation of fermions from the event horizon of the dynamic Kerr black hole is modified in this paper by considering the correction of dispersion relation.
\par
The Lorentz dispersion relation is considered to be one of the basic relations in modern physics, and related to the correlative theory research of general relativity and quantum field theory. Research on the quantum gravity theory have shown that the Lorentz relationship may be modified in the high energy field. In the study of string theory and quantum gravity theory, a dispersion relation has been proposed \cite{32,33,34,35,36}
\begin{equation}\label{1-1}
P_0^2=\vec{P}^2+m^2-(LP_0)^\alpha \vec{P}^2.
\end{equation}
In the natural units, $P_0$ and $P$ denote the energy and momentum of particles, respectively. $m$ is the rest mass of particles, and $L$ is a constant of the Planck scale. $\alpha=1$ is used in the Liouville-string model \cite{34}. Kruglov got the modified Dirac equation in the case of $\alpha=2$ \cite{37}. The more general motion equation of fermions was proposed by Rarita and Schwinger, and called the Rarita-schwinger equation \cite{38}. According to (\ref{1-1}), we select $\alpha=2$, so the Rarita-schwinger equation in flat spacetime is given by
\begin{equation}\label{1-2}
\big(\bar\gamma^\mu\partial_\mu+\frac{m}{\hbar}-\sigma\hbar\bar\gamma^t\partial_t\bar\gamma^j\partial_j\big)\psi_{\alpha_1\cdots\alpha_k}=0,
\end{equation}
where $\bar\gamma^\mu$ is the gamma matrix in flat spacetime. $j$ and $\mu$ denote space and time coordinates, respectively. According to the relationship between the covariant derivative of the curved spacetime and the derivative of the flat spacetime, the Rarita-Schwinger equation in the curved spacetime of non-stationary symmetric Kerr black hole can be expressed as
\begin{equation}\label{1-3}
\big(\gamma^\mu D_\mu+\frac{m}{\hbar}-\sigma\hbar\gamma^\mathsf{v}D_\mathsf{v}\gamma^jD_j\big)\psi_{\alpha_1\cdots\alpha_k}=0,
\end{equation}
where $\gamma^\mu$ is the gamma matrix in the curved spacetime, and $\mathsf{v}$ denotes the advanced Eddington coordinate. In Eq.(\ref{1-3}), when $k=0$, we have $\phi_{\alpha_1\wedge\alpha_k}=\phi$, in which case Eq.(\ref{1-3}) represents the Dirac equation of a spin of $1/2$; when $k=1$, Eq.(\ref{1-3}) describes the motion equation of the fermions with the spin of $3/2$. However, when $m=0$, the fermions with the spin of $3/2$ describe the Gravitino particle in the supersymmetry and supergravity theory, which are a kind of fermions associated with the graviton, and the study on such particles are likely to promote the development of quantum gravity theory.
\par
It is worth noting that, Eq.(\ref{1-3}) satisfy the following conditions
\begin{equation}\label{4}
\gamma^\mu\gamma^\nu+\gamma^\nu\gamma^\mu=2g^{\mu\nu}I,
\end{equation}
\begin{equation}\label{5}
\gamma^\mu\psi_{\mu\alpha_2\cdots\alpha_k}=D_\mu\psi^\mu_{\alpha_2\cdots\alpha_k}=\psi^\mu_{\mu\alpha_3\cdots\alpha_k},
\end{equation}
where $I$ is the unit matrix. In Eq.(\ref{1-3}), $D_\mu$ is defined as
\begin{equation}\label{6}
D_\mu\equiv \partial_\mu+\Omega_\mu,
\end{equation}
where $\Omega_\mu$ is the spin connection. In Eq.(\ref{1-3}), coupling constant $\sigma\ll1$, and $\sigma\hbar\gamma^\mathsf{v}D_\mathsf{v}\gamma^jD_j$ is very small quantity.
\par
In order to study the tunneling radiation of fermions in non-stationary curved spacetime, $S$ is used to represent the action function of particles, and the wave function of fermions is written as
\begin{equation}\label{7}
\psi_{\alpha_1\cdots\alpha_k}=\xi_{\alpha_1\cdots\alpha_k}e^{\frac{i}{\hbar}S}.
\end{equation}
For non-stationary and axisymmetric curved spacetime, there must be
\begin{equation}\label{8}
\partial_\varphi S=n,
\end{equation}
where $n$ is the angular momentum parameter of particles' tunneling radiation, and a constant for non-stationary axisymmetric black holes. Substituting (\ref{6})(\ref{7}) and (\ref{8}) into Eq.(\ref{1-3}), $\hbar$ is considered as a small quantity, and the lowest order is retained, so we obtain
\begin{equation}\label{9}
i\gamma^\mu\partial_\mu S\xi_{\alpha_1\cdots\alpha_k}+m\xi_{\alpha_1\cdots\alpha_k}+\sigma\partial_\mathsf{v}S\gamma^\mathsf{v}\gamma^j\partial_jS\xi_{\alpha_1\cdots\alpha_k}=0.
\end{equation}
Because of $\gamma^\mu\partial_\mu S=\gamma^\mathsf{v}\partial_\mathsf{v}S+\gamma^j\partial_jS$, the Eq.(\ref{9}) is abbreviated to
\begin{equation}\label{10}
\Gamma^\mu\partial_\mu S\xi_{\alpha_1\cdots\alpha_k}+M_D\xi_{\alpha_1\cdots\alpha_k}=0,
\end{equation}
where
\begin{eqnarray}\label{11}
\Gamma^\mu&=&\gamma^\mu-i\sigma\partial_\mathsf{v}S\gamma^\mathsf{v}\gamma^\mu, \\
M_D&=&m-\sigma(\partial_\mathsf{v}S)^2g^{\mathsf{v}\mathsf{v}}.
\end{eqnarray}
We use pre-multiplication $\Gamma^\nu\partial_\nu S$  to Eq.(\ref{10}), and get
\begin{equation}\label{13}
\Gamma^\nu\partial_\nu S\Gamma^\mu\partial_\mu S\xi_{\alpha_1\cdots\alpha_k}+M_D\Gamma^\nu\partial_\nu S\xi_{\alpha_1\cdots\alpha_k}=0,
\end{equation}
where
\begin{equation}\label{14}
\Gamma^\nu\Gamma^\mu=\gamma^\nu\gamma^\mu-2i\sigma\partial_\beta S\gamma^\beta\gamma^\nu\gamma^\mu+\mathcal{O}(\sigma^2).
\end{equation}
Now, after exchanging $\mu$ and $\nu$ in Eq.(\ref{13})and comparing them with Eq.(\ref{13}), we get
\begin{eqnarray}\label{15}
&&\big[\frac{\gamma^\mu\gamma^\nu+\gamma^\nu\gamma^\mu}{2}\partial_\mu S\partial_\nu S+m^2-2\sigma mg^{\mathsf{v}\mathsf{v}}(\partial_\mathsf{v}S)^2-2i\sigma\partial_\mathsf{v}Sg^{\mathsf{v}\beta}\partial_\beta S\gamma^\mu\partial_\mu S\big]\xi_{\alpha_1\cdots\alpha_k}+\mathcal{O}(\sigma^2) \nonumber\\
&&=\big[g^{\mu\nu}\partial_\mu S\partial_\nu S+m^2-2\sigma mg^{\mathsf{v}\mathsf{v}}(\partial_\mathsf{v}S)^2-2i\sigma\partial_\mathsf{v}Sg^{\mathsf{v}\beta}\partial_\beta S\gamma^\mu\partial_\mu S\big]\xi_{\alpha_1\cdots\alpha_k}+\mathcal{O}(\sigma^2) \nonumber\\
&&=0.
\end{eqnarray}
Eq.(\ref{15}) is further simplified to
\begin{equation}\label{16}
i\sigma\gamma^\mu\partial_\mu S\xi_{\alpha_1\cdots\alpha_k}+M_\xi\xi_{\alpha_1\cdots\alpha_k}=0,
\end{equation}
where
\begin{equation}\label{17}
M_\xi=\frac{g^{\mu\nu}\partial_\mu S\partial_\nu S+m^2-2\sigma mg^{\mathsf{v}\mathsf{v}}(\partial_\mathsf{v}S)^2}{2\partial_\mathsf{v}Sg^{\mathsf{v}\beta}\partial_\beta S}.
\end{equation}
We use pre-multiplication $-i\sigma\gamma^\mu\partial_\mu S$  to Eq.(\ref{16}),and exchange $\mu$ and $\nu$. Then, we add it to (\ref{16}) and divide by $2$, we get
\begin{equation}\label{18}
\big[\sigma^2g^{\mu\nu}\partial_\mu S\partial_\nu S+M_\xi^2\big]\xi_{\alpha_1\cdots\alpha_k}=0.
\end{equation}
This is a matrix equation, actually an eigen matrix equation. The condition that the equation has a non-trivial solution is that the value of the determinant corresponding to its matrix is $0$, that is
\begin{equation}\label{19}
\sigma^2g^{\mu\nu}\partial_\mu S\partial_\nu S+M_\xi^2=0.
\end{equation}
Ignoring $\mathcal{O}(\sigma^2)$, the modified Hamilton-Jacobi equation can be obtained from the above equation as
\begin{equation}\label{20}
g^{\mu\nu}\partial_\mu S\partial_\nu S+m^2-2\sigma mg^{\mathsf{v}\mathsf{v}}(\partial_\mathsf{v}S)^2=0.
\end{equation}
Obviously, the modified Hamilton-Jacobi equation (\ref{20}) is entirely different from the previously well-known Hamilton-Jacobi equation, with the addition of the modified term $2\sigma mg^{\mathsf{v}\mathsf{v}}(\partial_\mathsf{v}S)^2$. The equation (\ref{20}), derived from the modified Rarita-Schwinger equation, is not affected by the specific spin, and can describe the motion equation of any fermions in the semiclassical approximation method. For any fermions in non-stationary curved spacetime, it is convenient to study and modify characteristics of quantum tunneling radiation of fermions as long as the the properties of the curved spacetime and the action $S$ of fermions are known.
%
%
%%%%%%%%%%%%%%%%%%%%%%%%%%%%%%%%%
\section{Fermions' tunneling radiation of in non-stationary symmetric Kerr black hole}
\label{3}
%%%%%%%%%%%%%%%%%%%%%%%%%%%%%%%%%
%
%
%
\par\noindent
In the advanced Eddington coordinate, the line element of non-stationary symmetric Kerr black hole is expressed as
\begin{eqnarray}\label{21}
&&ds^2=-\big(1-\frac{2Mr}{\rho^2}\big)dr^2+2d\mathsf{v}dr-2\frac{2Mra\sin^2\theta}{\rho^2}d\mathsf{v}d\varphi-2a\sin^2\theta drd\varphi \nonumber\\
&&+\rho^2d\theta^2+\big[(r^2+a^2)+\frac{2Mra^2\sin^2\theta}{\rho^2}\big]\sin^2\theta d\varphi^2,
\end{eqnarray}
where $\rho^2=r^2+a^2\cos^2\theta$, $M=M(\mathsf{v})$, $a=a(\mathsf{v})$. According to (\ref{21}), the inverse tensors metric of the black hole is
\begin{equation}\label{23}
g^{\mu\nu}=\left(
  \begin{array}{cccc}
  g^{00} &~~ g^{01} &~~ 0 &~~ g^{03} \\
  g^{10} &~~ g^{11} &~~ 0 &~~ g^{13} \\
  0 &~~ 0 &~~ g^{22} & ~~0 \\
  g^{30} &~~ g^{31} &~~ 0 &~~ g^{33} \\
  \end{array}
\right),
\end{equation}
where
\begin{eqnarray}\label{24}
&&g^{00}=\frac{a^2\sin^2\theta}{\rho^2},~~g^{01}=g^{10}=\frac{r^2+a^2}{\rho^2}, \\ \nonumber
&&g^{03}=g^{30}=\frac{a}{\rho^2},~~g^{13}=g^{31}=\frac{a}{\rho^2},~~g^{11}=\frac{\Delta}{\rho^2}, \\ \nonumber
&&g^{22}=\frac{1}{\rho^2},~~g^{33}=\frac{1}{\rho^2\sin^2\theta},~~\Delta=r^2+q^2-2Mr.
\end{eqnarray}
According to (\ref{21}), the null hypersurface equation of the black hole is given by
\begin{equation}\label{25}
g^{\mu\nu}\frac{\partial f}{\partial x^\mu}\frac{\partial f}{\partial x^\nu}=0,
\end{equation}
Substituting (\ref{24}) into Eq.(\ref{25}), the equation at the event horizon of the black hole is expressed as
\begin{equation}\label{26}
a^2r_H^2\sin^2\theta+r_H^2-2mr_H+a^2+r'^2_H-2(r_H^2+a^2)\dot{r}_H=0,
\end{equation}
From (\ref{26}), we have
\begin{equation}\label{27}
r_H=\frac{M+[m^2-(1-2\dot{r}_H)(a^2+a^2\dot{r}_H\sin^2\theta+r'^2_H-2a^2\dot{r}_H)]^{1/2}}{1-2\dot{r}_H}.
\end{equation}
Obviously, the event horizon of the black hole $r_H$ is associated with $a(\mathsf{v})$, $M(\mathsf{v})$, $\partial_{\mathsf{v}}r|_{r=r_H}=\dot{r}_H$ and $\partial_\theta r|_{r=r_H}=r'_H$. Once we know the characteristics of the event horizon of the black hole, we can study the quantum tunneling radiation at the event horizon.
\par
The motion equation of fermions is given by the matrix equation (\ref{1-3}). From the above research, we can conclude that the motion equation of any half-integer fermions can be reduced to the equation (\ref{20}), and the equation (\ref{20}) is the modified Hamilton-Jacobi equation, where $S$ is the main function of Hamilton, also known as the action of fermions. Substituting (\ref{24}) and (\ref{8}) into Eq.(\ref{20}), the motion equation of the half-integer fermions in spacetime of the black hole is obtained
\begin{eqnarray}\label{28}
&&g^{00}(\frac{\partial S}{\partial\mathsf{v}})^2+2\frac{\partial S}{\partial\mathsf{v}}\frac{\partial S}{\partial r}+2g^{03}n\frac{\partial S}{\partial\mathsf{v}}+2g^{13}n\frac{\partial S}{\partial r}+g^{11}(\frac{\partial S}{\partial r})^2\\ \nonumber
&&+g^{22}(\frac{\partial S}{\partial\theta})^2+g^{33}n^2+m^2-2\sigma mg^{00}(\frac{\partial S}{\partial\mathsf{v}})^2=0.
\end{eqnarray}
Since the spacetime of the black hole is axially symmetric, $n$ is a constant according to $n=\partial S/\partial\varphi$. The equation (\ref{28}) is the motion equation of fermions in non-stationary Kerr black hole. Actually, the equation (\ref{28}) is a modified Hamilton-Jacobi equation in the non-stationary curved spacetime, where $S=S(\mathsf{v},r,\theta)$. In order to solve the equation, we need to use the general tortoise coordinate transformation as following
\begin{eqnarray}\label{29}
r_\star&=&r+\frac{1}{2\kappa}[r-r_H(\mathsf{v}_0,\theta_0)],\nonumber\\
\mathsf{v}_\star&=&\mathsf{v}-\mathsf{v}_0, \nonumber \\
\theta_\star&=&\theta-\theta_0,
\end{eqnarray}
According to (\ref{29}), we have
\begin{eqnarray}\label{30}
\frac{\partial}{\partial r}&=&\frac{2\kappa(r-r_H)+r_H}{2\kappa(r-r_H)}\frac{\partial}{\partial r_\star},\nonumber\\
\frac{\partial}{\partial\theta}&=&\frac{\partial}{\partial\theta_\star}-\frac{r'_H r_H}{2\kappa(r-r_H)}\frac{\partial}{\partial r_\star}, \nonumber \\
\frac{\partial}{\partial\mathsf{v}}&=&\frac{\partial}{\partial\mathsf{v}_\star}-\frac{\dot{r}_H r_H}{2\kappa(r-r_H)}\frac{\partial}{\partial r_\star},
\end{eqnarray}
Substituting (\ref{29}) and (\ref{30}) into (\ref{28}), and noticing
\begin{eqnarray}\label{31}
&&S=S(\mathsf{v}_\star,r_\star,\theta_\star),\nonumber\\
&&\frac{\partial S}{\partial\mathsf{v}_\star}=-\omega, \nonumber \\
&&\frac{\partial S}{\partial\theta_\star}=p_\theta,
\end{eqnarray}
where $\omega$ denotes the energy of fermions' tunneling radiation, $p_\theta$ is $\theta$ component of the generalized momentum of fermions, $\sigma$ is a small quantity, and $\sigma\dot{r}^2$ also a small quantity. The equation at the horizon of the black hole can be written as
\begin{eqnarray}\label{32}
\frac{A}{D}(\frac{\partial S}{\partial r_\star})^2+2\frac{\partial S}{\partial\mathsf{v}_\star}\frac{\partial S}{\partial r_\star}+\frac{B}{D}\frac{\partial}{\partial r_\star}+2\kappa(r-r_H)\frac{C}{D}=0
\end{eqnarray}
where
\begin{eqnarray}\label{33}
A&=&\frac{1}{2\kappa(r-r_H)}\Big\{g^{00}\dot{r}_H+2\dot{r}_H\big[2\kappa(r-r_H)+1\big]\\ \nonumber
&+&g^{11}\big[2\kappa(r-r_H)+1\big]^2+g^{11}\big[2\kappa(r-r_H)+1\big]^2+g^{22}{r'^2_H}\Big\}, \\
B&=&g^{22}p_\theta r'_H-g^{13}n-g^{03}n\dot{r}_H,\\
C&=&g^{00}\omega^2+g^{22}p_\theta^2+g^{33}n+m^2-2\sigma mg^{00}\omega^2,\\
D&=&g^{00}\dot{r}_H-g^{01}+2\sigma mg^{00}\dot{r}_H.
\end{eqnarray}
When $r\rightarrow r_H$, we have
\begin{equation}\label{37}
A(\frac{\partial S}{\partial r_\star})^2+2(\omega-\omega_0)\frac{\partial S}{\partial r_\star}=0,
\end{equation}
where
\begin{eqnarray}\label{38}
&&A|_{r\rightarrow r_H}\\ \nonumber
&&=\lim_{\substack{r\rightarrow r_H\\ \mathsf{v}\rightarrow\mathsf{v}_0\\ \theta\rightarrow\theta_0}} \frac{g^{00}\dot{r}_H^2-2g^{01}\dot{r}_H[2\kappa(r-r_H)+1]+g^{11}[2\kappa(r-r_H)+1]^2+g^{22}r'^2_H}{2\kappa(r-r_H)(g^{00}\dot{r}_H-g^{01}+2\sigma mg^{00}\dot{r}_H)}\\ \nonumber
&&=1.
\end{eqnarray}
From $A|_{r\rightarrow r_H}=1$, we get
\begin{eqnarray}\label{39}
\omega_0|_{r\rightarrow r_H}&=&\frac{g^{03}n\dot{r}_H+g^{22}p_\theta\dot{r}_H-g^{13}n}{g^{00}\dot{r}_H-g^{01}+2\sigma mg^{00}\dot{r}_H}\\ \nonumber
&=&\frac{an\dot{r}_H+p_\theta\dot{r}_H-an}{a^2\sin^2\theta\dot{r}_H-r_H^2-a^2+2\sigma m\dot{r}_Ha^2\sin^2\theta} \\ \nonumber
&=&\frac{an\dot{r}_H+p_\theta\dot{r}_H-an}{a^2\sin^2\theta\dot{r}_H-r_H^2-a^2}\Big(1-\frac{2\sigma m\dot{r}_Ha^2\sin^2\theta}{a^2\sin^2\theta\dot{r}_H-r_H^2-a^2}+\mathcal{O}(\sigma^2)\Big).
\end{eqnarray}
So the event horizon surface gravity is given by
\begin{eqnarray}\label{40}
\kappa&=&\frac{(1-2\dot{r}_H)r_H-M}{a^2\sin^2\theta_0\dot{r}_H(1+2\sigma m)-(r_H^2+a^2)(1-2\dot{r}_H)+4Mr_H-(r_H^2+a^2)} \nonumber\\
&=&\frac{(1-2\dot{r}_H)r_H-M}{a^2\sin^2\theta_0\dot{r}_H(1+2\sigma m)-(r_H^2+a^2)(1-2\dot{r}_H)+4Mr_H}
\end{eqnarray}
We notice $r_H^2(1-2\dot{r}_H)-2Mr_H+a^2(1-2\dot{r}_H+\dot{r}_H^2\sin^2\theta_0)+r'^2_H=0$, and get
\begin{eqnarray} \label{41}
\kappa&=&\frac{(1-2\dot{r}_H)r_H-M}{2Mr_H-(1-\dot{r}_H)\dot{r}_Ha^2\sin^2\theta_0+r'^2_H+2\sigma ma^2\sin^2\theta_0} \\ \nonumber
&=&\frac{(1-2\dot{r}_H)r_H-M}{(1-2\dot{r}_H)[r_H^2+a^2(1-\dot{r}_H\sin^2\theta_0)]+r'^2_H+2\sigma ma^2\sin^2\theta_0} \\ \nonumber
&=&\frac{(1-2\dot{r}_H)r_H-M}{(1-2\dot{r}_H)[r_H^2+a^2(1-\dot{r}_H\sin^2\theta_0)]+r'^2_H} \\ \nonumber
&\times&\{1-\frac{2\sigma ma^2\sin^2\theta_0}{(1-2\dot{r}_H)[r_H^2+a^2(1-\dot{r}_H\sin^2\theta_0)]+r'^2_H}+\mathcal{O}(\sigma^2)\}
\end{eqnarray}
Obviously, the event horizon surface gravity is modified, and the modified term depends on $\theta_0$. It means that the correction is made in different angle directions. Due to $\frac{\partial S}{\partial r}=[1+\frac{1}{2\kappa(r-r_H)}]\frac{\partial S}{\partial r_\star}$, we have
\begin{equation}\label{42}
S=\frac{i\pi}{2\kappa}[(\omega-\omega_0)\pm(\omega-\omega_0)].
\end{equation}
Thus, the imaginary part of the total action and the quantum tunneling rate respectively are
\begin{equation}\label{43}
ImS_+-ImS_-=\frac{i\pi}{2\kappa}\big[(\omega-\omega_0)\pm(\omega-\omega_0)\big],
\end{equation}
\begin{equation}\label{44}
\Gamma=\Gamma_{emission}/\Gamma_{absption}=\exp[\frac{2\pi}{\kappa}(\omega-\omega_0)]
\end{equation}
Here, as shown in Eq.(\ref{41}), it is clear that $\kappa$ mentioned above is the event horizon surface gravity of the black hole. So, the event horizon temperature of the black hole is given by
\begin{equation}\label{45}
T|_{r=r_H}=\frac{\kappa}{2\pi}.
\end{equation}
It is worth noting that the temperature (\ref{45}) is the modified Hawking temperature, since $\kappa$ in Eq.(\ref{45}) is the modified surface gravity related to the correction term $2\sigma ma^2\sin^2\theta_0$. Obviously, the correction of tunneling rate, surface gravity and Hawking temperature at the event horizon of the black hole are not only related to the rates of the event horizon change $\dot{r}_H$, $r'_H$ and $M(\mathsf{v})$ of the black hole, but also to the correction of the angle parameter $\theta_0$.
%
%
%%%%%%%%%%%%%%
\section{Discussion}
\label{4}
%%%%%%%%%%%%%%%%%%%%
%
%
%
\par\noindent
In this paper, we study the quantum tunneling radiation of fermions in non-stationary curved spacetime by combining the modified Lorentz dispersion relation, and obtain the modified character of quantum tunneling radiation related to the effects of the Planck scale. The modified Dirac equation proposed by Kruglov is first extended to the modified Rarita-Schwinger equation for the more general fermions, and the modified Hamilton-Jacobi equation of fermions is obtained in the semiclassical approximation method. Then, we study the quantum tunneling radiation of fermions in curved spacetime of non-stationary symmetric Kerr black hole using the modified Hamilton-Jacobi equation, and obtain the correction of Hawking temperature and tunneling rate of fermions. Interestingly, we found that the modified Hawking temperature at the event horizon of the black hole depends not only on the rates of the event horizon change $\dot{r}_H$, $r'_H$ and $M(\mathsf{v})$ of the black hole, but also on the correction of the angle parameter $\theta_0$. It means that the correction of Hawking radiation is not only the radial property of the black hole, but also related to the angular property of the black hole.
\par
In the study of quantum tunneling radiation of black holes, people first modified Hawking pure thermal radiation, and then modified character of tunneling radiation from stationary black holes. With the research on quantum gravity effect, we combine the deformed dispersion relation to modify the tunneling radiation of non-stationary symmetric Kerr black hole effectively. We believe that the correction of tunneling radiation from other types of curved spacetime will yield some interesting results. This paper only provides a method to modify quantum tunneling radiation, and further research is needed.
%%%%%%%%%%%%%%%%%%%%%%%%%%%
\section{Acknowledgements}
\par\noindent
This work is supported by the National Natural Science Foundation of China (Grant No. 11573022, No. 11805166), and by the starting funds of China West Normal University with Grant No.17YC513 and No.17C050.
%%%%%%%%%%%%%%%%%%%%%%%%%%%
%
%
%%%%%%%%%%%%%%%%%

%%%%%%%%%%%%
%
%
%%%%%%%%%%%%%%%%%%%

\begin{thebibliography}{99}
%%%%%%%%%%%%%%%%
%
%
\bibitem{1}
S. W. Hawking, Nature \textbf{248}, (1974) 30.

\bibitem{2}
S. W. Hawking, Commum. Math. Phys. \textbf{43}, (1975) 199.

\bibitem{3}
S. P. Robinson, F. Wilczek, Phys. Rev. Lett. \textbf{95}, (2005) 011303.

\bibitem{4}
T. Damoar, R. Ruffini, Phys. Rev. D \textbf{14}, (1976) 332.

\bibitem{5}
S. Sannan, Gen. Relativ. Gravit. \textbf{20}, (1988) 239.

\bibitem{6}
P. Kraus, F. Wilczek, Nucl. Phys. B \textbf{433}, (1995) 403.

\bibitem{7}
M. K. Parikh, F. Wilczek, Phys. Rev. Lett. \textbf{85}, (2000) 5042.

\bibitem{8}
M. K. Parikh, arXiv:hep-th/0402166.

\bibitem{9}
S. Hemming, E. Keski-Vakkuri, Phys. Rev. D \textbf{64}, (2001) 044006.

\bibitem{10}
Q. Q. Jiang, S. Q. Wu, X. Cai, Phys. Rev. D \textbf{75}, (2007) 064029.

\bibitem{11}
S. Iso, H. Umetsu, F. Wilczek, Phys. Rev. D \textbf{74}, (2006) 044017.

\bibitem{12}
A. J. M. Medved,  Phys. Rev. D \textbf{66}, (2002) 124009.

\bibitem{13}
J. Y. Zhang, Z. Zhao, Phys. Lett. B \textbf{638}, (2006) 110.

\bibitem{14}
K. Srinivasan, T. Padmanabhan, Phys. Rev. D \textbf{60}, (1999) 24007;
S. Shankaranarayanan, T. Padmanabhan, K. Srinivasan, Class. Quantum Grav. \textbf{19}, (2002) 2671.

\bibitem{15}
R. Banerjee and B. R. Majhi, Phys. Lett. B \textbf{690}, (2010) 83;
R. Banerjee and B. R. Majhi, Phys. Lett. B \textbf{675}, (2009) 243;
R. Banerjee and B. R. Majhi, Phys. Lett. B \textbf{674}, (2009) 218;
R. Banerjee and S. K. Modak, JHEP \textbf{11} (2009) 073.

\bibitem{16}
R. Kerner, R. B. Mann, Class. Quantum Grav. \textbf{25}, (2008) 095014;
R. Kerner, R. B. Mann, Phys. Lett. B \textbf{665}, (2008) 277.

\bibitem{17}
K. Lin, S. Z. Yang, Phys. Rev. D \textbf{79}, (2009) 064035; Phys. Lett. B \textbf{674}, (2009) 127; Int. J. Theor. Phys. \textbf{48}, (2009) 2061; Chinese Phys. B \textbf{20}, (2011) 110403.

\bibitem{18}
S. K. Modak, Phys. Lett. B \textbf{671}, (2009) 167.

\bibitem{19}
S. Sarkar and D. Kothawala, Phys. Lett. B \textbf{659}, (2008) 683.

\bibitem{20}
S. W. Zhou and W. B. Liu, Phys. Rev. D \textbf{77}, (2008) 104021.

\bibitem{21}
J. Hu, H. Yu, JHEP \textbf{1209}, (2012) 062.

\bibitem{22}
C. K. Ding, M. J. Wang and J. L. Jing, Phys. Lett. B \textbf{676}, (2009) 99.

\bibitem{23}
P. Mitra, Phys. Lett. B \textbf{648}, (2007) 240.

\bibitem{24}
E. T. Akhmedov, V. Akhmedova, D. Singleton, Phys. Lett. B \textbf{642} (2006) 124.

\bibitem{25}
R. D. Criscienzo, M. Nadalini, L. Vanzo, S. Zerbini and G. Zoccatelli, Phys. Lett. B \textbf{657}, (2007) 107.

\bibitem{26}
S. H. Mehdipour, Phys. Rev. D \textbf{81}, (2010) 124049.

\bibitem{27}
M. A. Rahman and M. I. Hossain, Phys. Lett. B \textbf{712}, (2012) 1.

\bibitem{28}
G. P. Li, J. Pu, Q. Q. Jiang, X. T. Zu, Euro. Phys. J. C \textbf{77}, (2017) 666;
J. Pu, Y. Han, Int. J. Theor. Phys. \textbf{55}, (2016) 5077;
J. Pu, Y. Han, Int. J. Theor. Phys. \textbf{56}, (2017) 2485;
J. Pu, Y. Han, Int. J. Theor. Phys. \textbf{56}, (2016) 2061.

\bibitem{29}
Z. W. Feng, H. L. Li, S. Z. Yang, X. T. Zu, Euro. Phys. J. C \textbf{76}, (2016) 212.

\bibitem{30}
H. L. Li, Z. W. Feng, S. Z. Yang, X. T. Zu, Euro. Phys. J. C \textbf{78}, (2018) 768.

\bibitem{301}
X. X. Zeng, D. Y. Chen, L. F. Li, Phys. Rev. D \textbf{91}, (2015) 046005;
X. X. Zeng, X. M. Liu, W. B. Liu, Eur. Phys. J. C  \textbf{72}, (2012) 1967;
X. X. Zeng, W. B. Liu, arXiv:1106.0548;
X. X. Zeng, S. W. Zhou, W. B. Liu, Chin. Phys. B \textbf{21}, (2012) 090402.

\bibitem{302}
Y. W. Han, X. X. Zeng, Y. Hong, Eur. Phys. J. C  \textbf{79}, (2019) 252;
X. X. Zeng, L. F. Li, Adva. High Energy Phys. \textbf{2016}, (2016) 6153435;
X. M. Liu, H. B. Shao, X. X. Zeng, Adva. High Energy Phys. \textbf{2017}, (2017) 6402101.

\bibitem{303}
Q. Q. Jiang, S. Q. Wu, Phys. Lett. B \textbf{647}, (2007) 200;
Q. Q. Jiang, X. Cai, JHEP \textbf{1011}, (2010) 066;
Q. Q. Jiang, D. Y. Chen, D. Wen, JCAP \textbf{1311}, (2013) 027.

\bibitem{304}
S. Z. Yang, K. Lin. J. Li, Q. Q. Jiang, Adva. High Energy Phys. \textbf{2016}, (2016) 7058764.

\bibitem{305}
Q. Q. Jiang, Phys. Rev. D \textbf{78}, (2008) 044009; Phys. Lett. B \textbf{666}, (2008) 517; Eur. Phys. J. C \textbf{72}, (2012) 2012.

\bibitem{306}
S. Z. Yang, J. Chinese West Normal Univ.(Nat. Sci) \textbf{37}, (2016) 126.

\bibitem{31}
S. Z. Yang, K. Lin, SCIENTIA SINICA Physica, Mechanica and Astronomica \textbf{49}, (2019) 019503.

\bibitem{32}
G. Amelino-Camelia, Int. J. Mod. Phys. D \textbf{11}, (2002) 35;
G. Amelino-Camelia, New J. Phys. \textbf{6}, (2004) 188.

\bibitem{33}
J. Magueijo and L. Smolin, Phys. Rev. Lett. \textbf{88}, (2002) 190403;
J. Magueijo and L. Smolin, Phys. Rev. D \textbf{67}, (2003) 044017.

\bibitem{34}
J. R. Ellis, N. E. Mavromatos and D. V. Nanopoulos, Phys. Lett. B \textbf{293}, (1992) 37;
J. R. Ellis, N. E. Mavromatos and D. V. Nanopoulos, Chaos Solitons Fractals \textbf{10}, (1999) 345;
J. R. Ellis, N. E. Mavromatos and A. S. Sakharov, Astropart. Phys. \textbf{20}, (2004) 669.

\bibitem{35}
S. I. Kruglov, Mod. Phys. Lett. A \textbf{28}, (2013) 1350014.

\bibitem{36}
T. Jacobson, S. Liberati and D. Mattingly, Nature \textbf{424}, (2003) 1019.

\bibitem{37}
S. I. Kruglov, Phys. Lett. B \textbf{718}, (2012) 228.

\bibitem{38}
W. Rarita and J. Schwinger, Phys. Rev. \textbf{60}, (1941) 61.
%
%%%%%%%%%%%%
\end{thebibliography}
\end{document}